\documentclass[aps,prb,showpacs,twocolumn]{revtex4}
\usepackage{amsmath,amssymb,array}
\usepackage{graphicx}
\usepackage{subfigure}
\tighten
\newcommand{\beq}{\begin{eqnarray}}
\newcommand{\eeq}{\end{eqnarray}}
\renewcommand{\vec}[1]{{\mathbf{#1}}}
\begin{document}
\title
{
 Superfluid density near the critical temperature in the presence of random
 planar defects
 }
\author{D. Dalidovich, A.J. Berlinsky and C. Kallin}
\vspace{.05in}
%
\address
{Department of Physics and Astronomy, McMaster University,\\
Hamilton, Ontario, Canada L8S 4M1 }
\date{\today}
%

%
\begin{abstract}
The superfluid density near the superconducting transition is investigated
in the presence of spatial inhomogeneity in the critical temperature.
Disorder is accounted for by means of a random $T_c$ term
in the conventional Ginzburg-Landau action for the superconducting
order parameter. Focusing on the case where a low-density of randomly distributed
planar defects are responsible for the variation of $T_c$,
we derive the lowest order correction to the superfluid density
in powers of the defect concentration. The correction is calculated assuming
a broad Gaussian distribution for the strengths of the defect potentials.
Our results are in a qualitative agreement with the superfluid density
measurements in the underdoped regime of high-quality YBCO crystals by Broun and co-workers.
\end{abstract}

\maketitle

\section{Introduction}

The superconducting transition measured in real materials
is often smeared or broadened in temperature in a way that
correlates with sample quality or disorder.  A sharper
transition is taken as a signature of a higher quality
sample.  On the other hand, a straightforward application
of the Harris criterion\cite{Harris} implies that uncorrelated
disorder is irrelevant
and does not affect the nature of the superconducting
transition. The Harris criterion is modified for
correlated disorder and
implies that the transition can be broadened,
depending on the nature or dimension of the correlation.

In addition, to the situations considered by Harris,
the existence of rare regions (analogous to Lifshitz tails\cite{lifshitz}
in the density of states of a disordered semiconductor)
with higher than average critical temperatures may also significantly
affect the
properties close to $T_c$ leading to a smeared behavior of
the order parameter as the transition is approached from
the ordered phases.  Such rare regions can occur with or
without correlations of the disorder. This type of behavior
is expected to be especially pronounced for cases in which the bare
superconducting correlation length $\xi_0$ is short. High temperature
superconductors in the underdoped regime, having a short coherence
length at low temperatures, present a good example of systems with
increased sensitivity to various types of correlated disorder.
Furthermore, many of these materials exhibit twin boundaries,
grain boundaries and/or disorder due to oxygen chains, all
of which are examples of correlated disorder.  Therefore,
the high temperature superconductors present a good example
of systems with increased sensitivity to various types of
disorder.

High-$T_c$ superconductors, because they are unconventional
in the sense that the gap averages to zero over the Fermi surface,
are very sensitive to disorder, although the sensitivity to
uncorrelated random disorder is somewhat mitigated by
their short coherence lengths.\cite{Franz}
The role of disorder
in superconductors has been an important
subject of study for several decades\cite{AG,Radtke,Franz}.
Imperfections in the lattice structure are not only responsible
for the diffusive motion of charge carriers above $T_c$, but also lead to
non-uniformity of the attractive interaction between them,
ultimately giving rise to a spatial variation of the local critical
temperature \cite{LO,Ioffe}. In addition to structural irregularities,
inhomogeneities in $T_c$ may be caused by the coexistence of
superconductivity and various density
waves\cite{machida,balseiro,sach}. Indeed, the simultaneous treatment
of several competing orders may be at the heart of a complete description of
high-temperature superconductors\cite{kiv,carlson} and other strongly
correlated systems.

Thus, obtaining a comprehensive understanding of all aspects of the
complicated interplay between superconductivity and various types of
inhomogeneities, is definitely an important goal.
An experimentally relevant set of issues that one can specifically address
in this context, would include the influence of inhomogeneities on
the critical temperature, fluctuation corrections above the transition,
as well as the behavior of the order parameter just below $T_C$.
It is clear that if the characteristic length scale of
the inhomogeneity is large compared to the $T=0$ superconducting
correlation length, and the width of the distribution of critical temperatures
is small compared to some average $T_C$, the problem can be studied within
the framework of Ginzburg-Landau (GL) theory with a space-dependent
critical temperature $T_c (\vec r)$. The GL action is suitable for
describing universal properties of a system in the vicinity of
the critical point which are insensitive to the details of the microscopic
Hamiltonian.  However, the functional form of $T_c (\vec r)$ is non-universal
and is determined by the type of non-uniformity present in the system.

In this paper, we study the behavior of the superfluid density
near the transition in the presence of one specific type of
randomness in $T_c$, caused by the presence of widely
scattered planar defects. The reason for this choice is two-fold.
First, twinning planes, that may be present even in
high quality crystals, can be regarded as planar
defects locally increasing the critical temperature. If a twinning
plane is located at $x=0$, the Ginzburg-Landau (GL) action will
contain a $\delta$-function term $-u \delta (x) |\Psi (\vec r)|^2$
with $u>0$.\cite{khlyust}
Second, to study the effect of disorder in general, one needs to
find stable solutions of the GL equation for arbitrary $T_c (\vec r)$
and then average over realizations of disorder in expressions
for the physical quantities of interest. This is
difficult to accomplish because the GL equation is non-linear. The
problem becomes more tractable, however, if the correction to the
GL action responsible for the change in $T_c$ has the form mentioned
above, with $u \gtrless 0$.

In this work, we consider a special case in which the spatial
variation of the critical temperature is modeled by
randomly located planar defects at points $X_i$, each
contributing a term
\beq\label{imppot}
u (\vec r)|\Psi (\vec r)|^2 =-u \delta (x-X_i) |\Psi (\vec r)|^2
\eeq
to the GL action.
Assuming that these imperfections are dilute, we obtain a
correction to the uniform superfluid density that is
proportional to the defect concentration $n_i$.
The calculations are performed at the simplest Gaussian level of
GL theory. The disorder potential $u$
is taken to have a Gaussian distribution with zero average value. The
presence of defects that locally enhance the critical temperature, gives
rise to a spatially decaying solution for the order parameter around them.
The leads to a small but non-zero total transverse response at
temperatures greater than $T_c^{(0)}$, the critical temperature of the pure
superconductor. The actual critical temperature corresponds to the region
of strongest $T_c$-enhancement and enters as a non-universal parameter
in our treatment.

Our work is motivated by recent experiments in the underdoped regime
of high-quality YBCO crystals.\cite{Broun} Experiments, done using the
cavity perturbation method, show that, despite the high quality
of the samples, vestiges of
\newcommand{\cbu}{\color{blue}} finite superfluid density persist
even for temperatures above $T_c^{(0)}$.
Our findings, based on the assumption of dilute planar
inhomogeneities in the form of twin boundaries, qualitatively explain
the results of measurements close to the critical temperature.

\section{Method of expansion in powers of the defect concentration}
\label{sec2}

\begin{widetext}


The starting point of our calculations close to criticality is the
Ginzburg-Landau (GL) free energy in terms of the local superconducting
order parameter $\Psi ({\bf r})$:\cite{Schmidt}
\beq\label{GL}
{\cal F}={\cal F}_n+\int
d {\bf r} \left\{ \frac{1}{2m} |\vec \nabla \Psi ({\bf r})|^2
+ \alpha(T) |\Psi ({\bf r})|^2 + U({\bf r}) |\Psi ({\bf r})|^2 +
\frac{b}{2} |\Psi ({\bf r})|^4 \right\}.
\eeq
\end{widetext}
Here ${\cal F}_n$ is the free energy of the normal system and
\beq\label{al}
\alpha (T) = \displaystyle a \left( \frac{ T -T_c^{(0)} }{T_c^{(0)}} \right)
\eeq
is the distance from the critical temperature, $T_c^{(0)}$, of a homogeneous
superconductor.
We assume that the deviations of $T_c (\vec r)$ from  $T_c^{(0)}$,
described by $U({\bf r})$,
occur in regions of a size greater than or of order
the $T=0$ correlation length, $\xi_0 =\xi (T=0)$, but small compared to
the correlation length near $T_c^{(0)}$.
This assumption justifies the use of the GL formalism for both conventional
and short coherence length superconductors, provided
that they are not too far from the critical temperature. In this case,
if the regions where the critical temperature differs sufficiently
from $T_c^{(0)}$ are located around points $\vec R_i$, we can
quite generally model
the randomness in Eq.~(\ref{GL}) by
\beq\label{uz}
U(\vec r )=\sum_i u_i (\vec r -\vec R_i ).
\eeq
In the subsequent treatment, we will refer to these regions as defects.
The functions $u_i (\vec r -\vec R_i)$, that we will henceforth call
 the potentials, are presumed to be quickly
decaying with $|\vec r -\vec R_i|$.
From the above considerations, the characteristic lengths of this
decay must exceed $\xi_0$ but be small compared to the correlation
length $\xi (T)$ close enough to the critical temperature.
In addition, it must be much smaller than the average separation
between the positions of the defects $n_i^{-1/d}$.
In the Gaussian approximation of GL theory $\xi (T)=1/\sqrt{2m\alpha (T)}$.

The equilibrium distribution of the superconducting order parameter,
$\Psi_0 ({\bf r})$, follows from the solution of the saddle point
GL equation that is derived by varying Eq.~(\ref{GL}) with respect
to $\Psi^* ({\bf r})$:
\beq\label{sadpo}
 \left[ -\frac{\vec \nabla ^2}{2m} + \alpha(T) + U({\bf r}) \right]
\Psi_0 ({\bf r}) + b \left| \Psi_0({\bf r}) \right|^2 \Psi_0 ({\bf r}) =0.
\eeq
For a given disorder potential, the actual transition temperature, $T_c$,
is determined from the value of $\alpha (T_c)$ for which a non-zero
solution of $\Psi_0 ({\bf r})$ first appears.
This happens when the eigenvalue spectrum of the operator
\beq\label{l0U}
{\hat L}_{0} [U] = -\frac{\vec \nabla ^2}{2m} + \alpha(T) + U({\bf r})
\eeq
crosses zero. The ensuing distribution  $\Psi_0 ({\bf r})$
can be chosen real and positive everywhere in space and must be stable.
The stability conditions can be determined if one expands the
generally complex order parameter $\Psi ({\bf r})$
around $\Psi_0 ({\bf r})$  in Eq.~(\ref{GL}),
\beq
\Psi ({\bf r})=\Psi_0 ({\bf r})+ \Psi_{\parallel} ({\bf r})+
i\Psi_{\perp} ({\bf r}).\nonumber
\eeq
One can then easily infer that a non-negative
eigenspectrum for the operators
\beq\label{LUperp}
{\hat L}_{\perp} [U]= -\frac{\vec{\nabla}^2}{2m} + \alpha(T) + U({\bf r})
+b \Psi_0^2 ({\bf r}),
\eeq
\beq\label{LUpar}
{\hat L}_{\parallel} [U] = -\frac{\vec{\nabla}^2}{2m} + \alpha(T) + U({\bf r})
+3b \Psi_0^2 ({\bf r}),
\eeq
is a necessary condition for stability.
The problem of determining the transition point and finding the stable
solution below the transition analytically for the general
form, $ U({\bf r})$, is a daunting task. However, one can simplify
the problem if the concentration of defects $n_i$
is small. If one assumes that every defect
potential, $u_i (\vec r -\vec R_i)$, is characterized by the same set
of parameters $\{ u \}$, one can employ the method
of expansion in powers of concentration $n_i$.\cite{Lifshitz}

Consider the function
\beq
F_N ( \{ u_1 \}\vec R_1, \ldots ,\{ u_N \}\vec R_N ;\vec r) \nonumber
\eeq
that describes the spatial dependence of some quantity of interest
and is calculable based on the GL action in the presence of $N$ defects.
We presume also that every defect located at point $\vec R_i$ has a
potential characterized by the specific parameter set $\{ u_i \}$.
The values of the parameters  $\{ u \}$ are distributed
according to the distribution $\mathcal{P}(\{ u \})$,
satisfying the normalization condition
\beq\label{unorm}
\int  \mathcal{P} (\{ u \}) {\cal D}\{ u \}  =1.
\eeq
In Eq.~(\ref{unorm}), ${\cal D}\{ u \}$ formally means the integration
over all variables in the set $\{ u \}$. We are interested in the value
of the function $F_N ( \{ u_1 \}\vec R_1, \ldots ,\{ u_N \}\vec R_N ;\vec r)$
that is an average over positions $\vec R_i$ as well as
parameters $\{ u_i \}$. If one denotes the positional average by
angular brackets, the full average can be written in the form:
\begin{widetext}
\beq\label{avgen}
\overline{F_N ( \{ u_1 \}\vec R_1 \ldots \{ u_N \}\vec R_N ;\vec r) }=
\int   \mathcal{P} (\{ u_1 \}) \ldots   \mathcal{P} (\{ u_N \})
\mathcal{D}  \{ u_1 \}\ldots   \mathcal{D} \{ u_N \}
\bigl\langle F_N ( \{ u_1 \}\vec R_1, \ldots ,\{ u_N \}\vec R_N ;\vec r)
\bigr\rangle.
\eeq
Regarding the concentration of defects as small, one can
formally write the average Eq.(\ref{avgen}) in the form of a
series in powers of $n_i$. The details of the corresponding
derivation are presented in Appendix A. As follows from Eqs.~(\ref{genfor})-
(\ref{spav}), up to the first order in $n_i$,
\beq\label{sav}
\bigl\langle F_N ( \{ u_1 \} \vec R_1, \ldots ,\{ u_N \} \vec R_N ;\vec r )
\bigr\rangle = F_0 (\vec r) +
n_i \int \left[ F_{1} (\{ u_1 \};\vec R_1;\vec r) - F_0 (\vec r)
\right] d\vec R_1,
\eeq
so that
\beq\label{spav1}
\overline{F_N ( \{ u_1 \}\vec R_1 \ldots \{ u_N \}\vec R_N ;\vec r) }=
F_0 (\vec r) + n_i \int \mathcal{P} (\{ u \}) {\cal D}\{ u \}
\int \left[ F_{1} (\{ u \};\vec R;\vec r) - F_0 (\vec r)
\right] d\vec R
\eeq
Eq. (\ref{spav1}) contains the lowest order correction
to the function $F_0 (\vec r)$, the quantity of interest in the
absence of any defects. The calculation of this correction requires
the knowledge of function $F_{1} (\{ u \};\vec R;\vec r)$ --
the quantity of interest in the presence of just one defect
located at point $\vec R$. We should mention that the approach
described in Appendix A, provides a way to reduce the level of
complexity of the initial problem, since it reduces to calculations
in the presence of just a finite number of defects.
This task is simpler, although in practice one has to limit the
treatment to the level of one or at most two lowest orders in $n_i$.
An important assumption made in the development of this approach
is that all integrations in Eq.~(\ref{spav}),
in every term of expansion in $n_i$, do not lead to divergences.
Convergence must be maintained for all parameters in the set $\{ u \}$
and all values of $\alpha (T)$ especially the point $\alpha (T)=0$.
This property, ensuring that this method of expansion is controlled,
is far from being a forgone conclusion, and must be carefully addressed
once the specific form of the defect potential is chosen.
As will be shown below, for the essentially one-dimensional potentials
such as those given by Eq.~(\ref{imppot}), the convergence is maintained
for all $\alpha (T)$ in the first order of expansion in $n_i$. Although,
we will be concerned below only with this lowest order, we believe that
the procedure is well-behaved at all orders, as long as defects are
parallel to each other
and the one-dimensional character of the problem is maintained.

\section{Superfluid density}

In this Section, we calculate the superfluid density in the presence
of randomly located planar defects based on the expansion in powers of
defect concentration $n_i$.
We will limit ourselves to calculations up to first order in $n_i$.
It is useful, however, to first discuss the qualitative behavior
of the superconducting order parameter in the presence of
defects, without specifying the dimensionality of the problem or the form
of the potential $u (\vec r -\vec R_i)$.
In the absence of randomness, the superfluid density is\cite{Schmidt}
\beq\label{cleanden}
\rho_s^{(0)} (T)= [\Psi_{0}^{(0)} (T)]^2 =\left\{
\begin{array}{cl}
0, & \quad \alpha (T)>0
\\
\left| \alpha (T) \right| /b,& \quad \alpha (T)<0;
\end{array}\right.
\eeq
To calculate the first order correction in $n_i$ to this result, one
needs to solve the saddle point equation for the order parameter
$\Psi_{0}^{(1)} (\vec R_i;\vec r)$ in the presence of one defect
located at point $\vec R_i$:
\beq\label{opimp}
 \left[ -\frac{\vec{\nabla}^2}{2m} + \alpha(T) + u(\vec r -\vec R_i) \right]
\Psi_0^{(1)} (\vec R_i;\vec r) + b \left| \Psi_0^{(1)} (\vec R_i;\vec r)
\right|^2 \Psi_0^{(1)} (\vec R_i;\vec r) =0.
\eeq
It is clear that
$\Psi_{0}^{(1)} (\vec R_i;\vec r)=\Psi_{0}^{(1)} (\vec r -\vec R_i)$,
and without loss of generality we can consider the defect to be located
at $\vec R_i =0$.
The non-zero real and positive solution of this equation occurs
at the point where the eigenvalue spectrum of the operator ${\hat L}_{0} [u]$,
containing a single-defect potential, reaches zero.
In analogy with Eq.~(\ref{l0U}),
\beq\label{L0u}
\hat{L}_{0} [u] = -\frac{\vec \nabla ^2}{2m} + \alpha(T) +
u(\vec r).
\eeq
Provided that the solution obtained from  Eq.~(\ref{opimp})
is stable, we can write down the general formula for the
superfluid density averaged over the randomness in $u(\vec r)$
\beq\label{rho}
\overline{\rho_s} (T) =\left\{
\begin{array}{cl}
        n_i \displaystyle\int_{{\mathcal C}}
 \Tilde{\mathcal{P}} (\{ u \}) \mathcal{D}  \{ u \} \int
[ \Psi_{0}^{(1)} (\vec r) ]^2 d\vec r
 , & \quad \alpha (T)>0;
\\
|\alpha (T)|/b + n_i \displaystyle \int_{{\mathcal C}}
\Tilde{\mathcal{P}} (\{ u \}) \mathcal{D}  \{ u \} \int
\bigl\{ [ \Psi_{0}^{(1)} (\vec r) ]^2
-|\alpha (T)|/b \bigr\} d\vec r
 ,& \quad \alpha (T)<0.
 \end{array}\right.
\eeq
\end{widetext}
Expression Eq.~(\ref{rho}) follows straightforwardly from
Eq.~(\ref{spav1}), applied to the square of the order parameter.
But the same result can be obtained using time-dependent Ginzburg-Landau
equation as a starting point.\cite{Schmidt}
One needs to calculate the transverse response, and
subsequently perform the average using the same method of expansion
in powers of $n_i$ in the long wavelength limit.\cite{Dalidovich}
The meaning of notations $\int_{{\mathcal C}}$ and
$\Tilde{\mathcal{P}} (\{ u \})$ in Eq.~(\ref{rho}) will be discussed below.
We only mention now that
we must carefully integrate, not over all possible values of parameters
from the set $\{ u \}$, but only over those realizations that, first,
give stable solutions for $\Psi_{0}^{(1)} (\vec r)$  and, second,
result in a defect-affected critical temperature not exceeding some fixed
value $T_c$.

If the system contains defects that give rise to a stable
positive non-zero solution of Eq.~(\ref{opimp}) at $T$ greater
than $T_c^{(0)}$, superconductivity must be
presumed shifted to higher temperatures.
In this case, a finite transverse response will be observed above
the critical temperature of the homogeneous sample.  To first order in $n_i$, the
actual transition point will be determined by the defect
that leads to the strongest enhancement
of $T_c$ in the sample.
This means that in experiments, the temperature $T_c$, at which the
superconducting response is first seen will be disorder-dependent and
non-universal. Very close to $T_c$, the superfluid density will be
tiny, since the order parameter will be determined by
contributions coming from a very small number of defects. But with
decreasing temperature, the fraction of defects giving rise to
non-zero solutions of Eq.~(\ref{opimp}) will increase, leading to
an increase in the superconducting response.
At $T>T_c^{(0)}$, the solutions for the order parameter,
$\Psi_{0}^{(1)} (\vec r -\vec R_i)$, will be localized
around the center of the defect at point $\vec R_i$.
Indeed, as long as $\alpha \equiv \alpha (T)>0$, in the absence of
any defects,
the only stable solution for the order parameter is zero. Hence,
it follows from Eq.~(\ref{opimp}) that for
functions $u(\vec r)$ that vanish quickly enough with distance,
\beq\label{psigr}
\Psi_{0}^{(1)} (\vec r ) = \frac{g_{>}(\vec r)}{\sqrt{b}}
 e^{-\sqrt{2m \alpha } r },
\eeq
when the condition
$\sqrt{m \alpha}r \gg 1$ is satisfied. The function
$g_{>} (\vec r)$ which has a weaker than exponential dependence on $r$,
is determined by the effective dimensionality of the problem.
For instance, in two dimensions $g_{>} (\vec r)=1/\sqrt{r}$
and does not contain any variables related to the potential.\cite{Millis}
For $\sqrt{m \alpha }r \le 1$, however, the functional form of the
decay of the order parameter is no longer exponential and depends
strongly on the non-universal characteristics of the function $u(\vec r)$.
Eq.~(\ref{psigr}) also describes the behavior of $\Psi_{0}^{(1)} (\vec r)$ at
all distances when $\alpha=0$ exactly.
The corresponding asymptotic forms are in fact written out in
Table 1 of Ref.~\onlinecite{Millis} and we will not
discuss them further here.

When $T<T_c^{(0)}$ and $\alpha <0$, the stable solution for the clean
system is given by $\Psi_{0}^{(0)} (T)=\sqrt{| \alpha | /b}$, and we
expect from Eq.~(\ref{opimp}), that finite values of $u(\vec r)$
will add some perturbation to this solution that falls off at infinity.
We look for solutions of the form
\beq\label{belowpsi}
\Psi_{0}^{(1)} (\vec r ) = \frac{1}{\sqrt{b}} \left[
\sqrt{|\alpha|}+ \psi (\vec r ) \right],
\eeq
where the real auxiliary function $\psi (\vec r )$ satisfies the
equation
\beq\label{forpsi}
\left[ -\frac{\vec{\nabla}^2}{2m} + 2|\alpha| + u(\vec r) \right]
\psi (\vec r) + \sqrt{|\alpha|} u(\vec r)\nonumber\\
+3\sqrt{|\alpha|} [ \psi (\vec r)]^2
+[ \psi (\vec r)]^3 =0.
\eeq
By analogy to Eq.~(\ref{psigr}) and provided that
$\sqrt{m|\alpha|} r  \gg 1$,, we can write
\beq\label{psile}
\psi^{(1)} (\vec r )=g_{<} (\vec r)
 e^{-2\sqrt{m |\alpha|} r },
\eeq
with the function $g_{<} (\vec r)$ having an asymptotic form similar
to that of $g_{>} (\vec r)$.
Again, closer to the defect when $\sqrt{m|\alpha|} r \sim 1$,
the crossover to a different
functional form, with stronger dependence on characteristics
of the potential, will take place.
Note also that the requirement for
$\Psi_{0}^{(1)} (\vec r)$ to be positive, does not prevent
the function $g_{<} (\vec r)$ from having both signs.
This  means that depending on the form and sign of
$u(\vec r )$, the order parameter may be either enhanced or suppressed
in the vicinity of a defect for $\alpha <0$.
Thus we conclude that the spatial variation of the order parameter
changes qualitatively when $\alpha$ passes through zero,
meaning that the defects play a different role in the system
above and below $T_c^{(0)}$. Above $T_c^{(0)}$,
only a portion of all defects will perturb the zero
value of $\Psi_{0}^{(0)}$ and, though widely scattered, they nevertheless
ensure a small but finite superconducting response. At the same time,
below $T_c^{(0)}$, every defect will affect the solution
$\Psi_{0}^{(0)}=\sqrt{|\alpha | /b}$,
but this just leads to a correction to the superfluid density
that becomes more and more innocuous with decreasing temperature.
In some sense, we can say that, because of these qualitative
differences, the point $T=T_c^{(0)}$ acts as a special kind of
critical point. Indeed, a simple inspection of Eq.~(\ref{rho}) reveals
that, despite the continuity of the superfluid density at
$\alpha =0$, its derivative with respect to temperature exhibits a jump.

It is then appropriate to ask what kind of a defect potential
$u(\vec r -\vec R_i)$ leads to an increase in the critical temperature.
To answer this question, consider the eigenvalue problem for
the operator ${\hat L}_{0} [u] -\alpha$,
where ${\hat L}_{0} [u]$ is given by Eq.~(\ref{L0u}):
\beq
\left[ -\frac{\vec{\nabla}^2}{2m} + u(\vec r-\vec R_i) \right]
\chi_{\epsilon} (\vec r-\vec R_i) = \epsilon \chi_{\epsilon} (\vec r-\vec R_i).
\eeq
This equation is nothing other than the Schr\"{o}dinger equation determining
the stationary states of a particle moving in the presence of
potential $u(\vec r-\vec R_i)$. For a potential which
falls off rapidly enough at infinity,
the spectrum of positive eigenvalues $\epsilon$
is continuous. It is describable by a number of quantum variables,
with $k=\sqrt{2m\epsilon}$ being one of them. The spectrum of negative
eigenvalues $E_n$, if any exist, is discrete.
All eigenvalues $\epsilon$ are explicit functions of all parameters
in the set $\{ u \}$.
It is easy to see then that the transition point, in the presence
of one defect, is determined by the smallest eigenvalue, $E_0$, of the operator
${\hat L}_{0} [u] -\alpha$.
From the condition $\alpha -|E_0| =0$ and Eq.~(\ref{al}) it follows
that, because of the defect, $T_c =T_c^{(0)}(1+|E_0|/a)$.
If there are no discrete levels, $E_0=0$ and no increase of the
critical temperature occurs. We note that
the points of instability coincide with the poles of the Green's function
of the operator ${\hat L}_{0} [u]$, obeying the following equation
\beq\label{gfLeq}
{\hat L_{0}}[u] {\cal G}^{(1)} (\vec R_i;\vec r,\vec r^{\prime})=
\delta (\vec r-\vec r^{\prime}).
\eeq
The superscript means that the Green's function are calculated
in the presence of only one defect located at $\vec R_i$. Above $T_c$,
\beq\label{gfL}
{\cal G}^{(1)} (\vec R_i;\vec r,\vec r^{\prime})=\sum_{\{ \epsilon \}}
\frac{\chi_{\epsilon} (\vec r -\vec R_i) \chi^{\star}_{\epsilon}
(\vec r^{\prime} -\vec R_i)}
{\alpha+\epsilon},
\eeq
where $\sum_{\{ \epsilon \}}$ formally denotes the summation and integration
over the discrete and continuous branches of the spectrum.
Since any defect breaks translational invariance,
${\cal G}^{(1)} (\vec R_i;\vec r,\vec r^{\prime})$
depends separately on $\vec r$ and $\vec r^{\prime}$ rather than
on $\vec r - \vec r^{\prime}$. The Green's function in the momentum
representation
${\cal G}^{(1)} (\vec R_i;\vec p,\vec p^{\prime})$
is the Fourier transform of Eq.~(\ref{gfL}). It
contains two momenta, $\vec p$ and $\vec p^{\prime}$, and
has the same pole structure but may be more straightforward to calculate
depending on the form of potential.

Following the same strategy that led to Eq.~(\ref{spav1}),
it is possible to write down the Green's function
averaged over the positions of defects and parameters of their
potentials.
In analogy with Eq.~(\ref{spav1}), up to first order in $n_i$:
\begin{widetext}
\beq\label{grspav1}
\overline{{\cal G}^{(N)} ( \{ u_1 \}\vec R_1 \ldots \{ u_N \}\vec R_N
;\vec r, \vec r^{\prime} ) } =
{\cal G}^{(0)} ( \vec r -\vec r^{\prime}) +
n_i \int \mathcal{P} (\{ u \}) {\cal D}\{ u \}
\int \left[  {\cal G}^{(1)} (\{ u \};\vec R; \vec r, \vec r^{\prime})
- {\cal G}^{(0)} ( \vec r -\vec r^{\prime}) \right] d\vec R
\eeq
\end{widetext}
An approximation that contains only the first power in $n_i$ is known as
a single-site approximation\cite{Elliott,Langer1,Langer2}. If all
defects have one and the same potential, the system is considered to have
binary disorder, in which case the integration over
${\cal D}\{ u \}$ would be absent in Eqs.(\ref{spav1}),(\ref{grspav1}).
For this widely studied simplified type of randomness, the Green's function
Eq.~(\ref{grspav1}) is obtainable as a result of resummation of a
certain class of diagrams\cite{Elliott, Schwabl}. If the distribution
of parameters characterizing the potentials $u(\vec r -\vec R_i)$
is broad enough, averaging over them with the weight $\mathcal{P} (\{ u \})$
plays an important role introducing an additional complicating
ingredient to the problem. The critical value of $\alpha (T)$ for which
the non-zero solution of Eq.~(\ref{opimp}) first appears in this case,
will properly coincide with the singularity in the disorder averaged
Green's function as can be seen from Eqs.~(\ref{gfL}),~(\ref{grspav1}).
It will be determined by the defect that induces the maximum local
$T_c$ in the sample.

Next we address the question of the stability of possible solutions
of Eq.~(\ref{opimp}). In the presence of one defect (presumed
located at $\vec R_i =0$),
the solution is stable if the eigenspectrum of operators
${\hat L}_{\perp} [u]$ and ${\hat L}_{\parallel} [u]$,
$\varepsilon_{\perp}$ and $\varepsilon_{\parallel}$,
written in analogy with Eqs.~(\ref{LUperp}) and~(\ref{LUpar}),
is non-negative. That is, one needs to analyze two equations:
\beq\label{Luperp}
\Bigl[ -\frac{\vec{\nabla}^2}{2m} + \alpha + u({\bf r})
+b [\Psi_0^{(1)} ({\bf r})]^2 \Bigr] f_{\perp} (\vec r) =
\varepsilon_{\perp} f_{\perp} (\vec r),
\eeq
\beq\label{Lupar}
\Bigl[ -\frac{\vec{\nabla}^2}{2m} + \alpha + u({\bf r})
+3b [\Psi_0^{(1)} ({\bf r})]^2 \Bigr] f_{\parallel} (\vec r) =
\varepsilon_{\parallel} f_{\parallel} (\vec r).
\eeq
To do this, we employ the following general mathematical result
for the spectra of second-order differential operators.
The eigenvalues can be ordered in a sequence of increasing
values, and the eigenfunction corresponding to the lowest eigenvalue
(the ground state eigenfunction) has no nodes
as a function of $\vec r$.\cite{Titchmarsh}
Eigenfunctions corresponding to higher energies must change sign somewhere
in space and are orthogonal to the ground state eigenfunction.
Comparing Eqs.~(\ref{opimp}) and~(\ref{Luperp}), we see that
the eigenfunction corresponding to $\varepsilon_{\perp }=0$
is given by $\Psi_0^{(1)} ({\bf r})$.
Hence if it is everywhere positive, we can claim that it is the
ground state of operator ${\hat L}_{\perp} [u]$. This result immediately
implies the conclusion that the lowest eigenvalue of Eq.~(\ref{Lupar}),
$\varepsilon_{\parallel 0}$, cannot be negative. Indeed, the ground state
eigenfunction $f_{\parallel 0} (\vec r)$ must be bounded and
can not change sign anywhere. Hence if we consider the ground states
of Eqs.~(\ref{Luperp}) and~(\ref{Lupar}), multiply
them respectively by $f_{\parallel 0} (\vec r)$ and
$\Psi_0^{(1)} ({\bf r})$, integrate over $d\vec r$ and then subtract
the first from the second, we find that
\beq
2b\int \left[\Psi_0^{(1)} ({\bf r}) \right]^3 f_{\parallel 0} (\vec r)
d\vec r = \varepsilon_{\parallel 0}  \int
\Psi_0^{(1)} ({\bf r}) f_{\parallel 0} (\vec r) d\vec r.\nonumber
\eeq
This immediately implies that $\varepsilon_{\parallel 0}>0$ is
the only possibility, and hence that all other
$\varepsilon_{\parallel}>0$ as well.
This result is just a simple manifestation
of the fact that, for predominantly positive potentials, the set of
eigenvalues shifts up.  We can not say, however, whether the ground
state belongs to the discrete spectrum or lies at the bottom
edge of the continuous one.
From Eq.~(\ref{Lupar}) it follows that for $\alpha>0$, the
continuous spectrum starts
at $\varepsilon_{\parallel}=\alpha$, while for $\alpha <0$,
$\varepsilon_{\parallel}=2|\alpha|$ is its lowest possible
eigenvalue. If $0<\varepsilon_{\parallel 0}<\alpha$ and $\alpha>0$
(or $2|\alpha|$ for $\alpha <0$ ), then the ground state eigenfunction
$f_{\parallel 0} (\vec r)$ belongs to the discrete branch of the spectrum and
falls off exponentially at infinity.
But if $\varepsilon_{\parallel 0}$ is the lowest possible
eigenvalue of the continuous spectrum, $f_{\parallel 0} (\vec r)$
tends to some non-zero constant as $|\vec r|\rightarrow\infty$.
Similar analysis of asymptotics following from Eq.~(\ref{Luperp})
leads to the result that for $\alpha <0$, the spectrum of
${\hat L}_{\perp} [u]$ is purely continuous and starts
right from the ground state zeroth eigenvalue.
If $\alpha >0$, however, one can not exclude the presence
of some additional energy levels belonging to the discrete
spectrum in the segment $0<\varepsilon_{\perp}<\alpha$.
To conclude, if one finds a solution of Eq.~(\ref{opimp})
$\Psi_0^{(1)} (\vec r)$ positive for all $\vec r$, it is guaranteed
to be stable. We are not aware of any general
analytic methods that allow us to solve
Eq.~(\ref{opimp}) because of the cubic non-linearity.
But the possibility that solutions may be found,
depending on the relation between $\alpha$ and the parameters of $u(\vec r)$,
seems quite realistic for smooth potentials of a given sign which decay
monotonically at infinity.

We are now in the position to discuss how to perform
the average over disorder realizations
$\int_{{\mathcal C}}$ in Eq.~(\ref{rho}),
and what limitations one should impose on the
distribution of randomness in order to obtain
physically sensible results for the superfluid density.
As discussed earlier, our approach implies that the actual critical
temperature is determined by the defect which
gives the greatest increase of $T_c^{(0)}$. This approximation is a
consequence of considering the problem in the lowest order in $n_i$
and suggests that, to compare the
theoretically calculated disorder-smeared behavior with
experimental data, the actual $T_c$ should be introduced by hand.
This can be achieved if the integral
$\int \mathcal{D}\{ u \}$ is performed,
not over all possible values from the set $\{ u \}$,
but rather over those of them that do not allow for
the defect-shifted critical temperatures greater than
the stipulated $T_c$. Subscript ${\mathcal C}$ in Eq.~(\ref{rho})
is used to indicate exactly that. Since in this case
$\int_{\mathcal C} \mathcal{P} (\{ u \}) \mathcal{D}\{ u \} <1$, it is
appropriate to introduce the normalized distribution,
\beq\label{disnew}
 \Tilde{{\mathcal P}} (\{ u \})=\frac{1}{{\mathcal A}}
\mathcal{P} (\{ u \}),
\quad
{\mathcal A}= \int_{\mathcal C} \mathcal{P} (\{ u \}) \mathcal{D}\{ u \}
\eeq
and employ it in the average over disorder potentials.

In addition, to eliminate the possibility of unphysical behavior of the
superfluid density as a function of temperature,
the disorder distribution must be regarded as symmetric about its
average. Namely, we must require the fulfillment of the constraint:
\beq\label{disconstr}
\int_{\mathcal C} \mathcal{P} (\{ u \}) \mathcal{D}\{ u \}
\int u(\vec r) d\vec r =0.
\eeq
To clarify its importance, we first note that
the positive sign of $\overline{\rho_s}$ does not follow
automatically from Eq.~(\ref{opimp}) when $\alpha <0$. Indeed,
if we divide Eq.~(\ref{opimp}) by
$b\Psi_{0}^{(1)} (\vec r -\vec R_i)$ and integrate over
space and $\mathcal{D} \{ u \}$, then using Eq. (\ref{rho}) for $\alpha <0$,
we find that
\beq
&\overline{\rho_s }=  \displaystyle\frac{|\alpha|}{b} + \nonumber\\
&\displaystyle\frac{n_i}{b}
 \int_{{\mathcal C}}
\mathcal{P} (\{ u \}) \mathcal{D}  \{ u \} \int
\left[
\frac{\nabla^2 \Psi_{0}^{(1)} (\vec r)}
{2m \Psi_{0}^{(1)} (\vec r)} -u(\vec r)
\right] d\vec r
\eeq
We can then integrate by parts the term containing $\Psi_{0}^{(1)} (\vec r)$,
with the help of Eqs.~(\ref{belowpsi}), (\ref{psile}).
Hence, if Eq.~(\ref{disconstr}) is satisfied, it follows that
\beq\label{rhobel}
\overline{\rho_s }=  \displaystyle\frac{|\alpha|}{b} +
\frac{n_i}{2m b} \int_{{\mathcal C}}
\mathcal{P} (\{ u \}) \mathcal{D}  \{ u \} \! \int
\left[
\frac{\nabla \Psi_{0}^{(1)} (\vec r)}{\Psi_{0}^{(1)} (\vec r)}
\right]^2  \! \! d\vec r
\eeq
The integrand in Eq.~(\ref{rhobel}) is always positive, ensuring that
$\overline{\rho_s } >0$ everywhere below $T_c^{(0)}$.
This could not be the case, had the left hand side of Eq.~(\ref{disconstr})
been negative. Although nothing wrong occurs
if it is positive, it is convenient  to ensure that
Eq.~(\ref{disconstr}) is satisfied by adjusting $T_c^{(0)}$
which so far has been assumed to be
the transition temperature of a disorder-free sample.

We now apply this general formalism to the case in which the defects
are described by the potential:
\beq\label{ourpot}
u (\vec r)=-u\delta (x-X_i)
\eeq
As mentioned in the Introduction, our model consists of a stack of
parallel planes that locally change the critical temperature.
The planes are infinite in $y$ and $z$ directions,
resulting in an essentially a one-dimensional problem.
Eq.~(\ref{ourpot}) also implies that the planes are formally
of zero thickness. Physically, this corresponds to a
situation in which the actual width of the planar defects
is of order $\xi_0$.
The only new parameter in the problem
having dimensions of length is
$1/(m|u|)$ which, together with the correlation length
$\xi=1/\sqrt{2m\alpha}$, determines the character of the solution
for the order parameter. Comparing those two length scales one can see
that any shift to $\alpha$ should be proportional to $mu^2$. The order
parameter $\Psi_{0}^{(1)} (\vec r)$ depends only on $x$, and
assuming, again without loss of generality, that the defect is at
the origin, we write the equation
\begin{widetext}
\beq\label{planimp}
\left[ -\frac{1}{2m} \frac{d^2}{dx^2}+ \alpha - u\delta (x) \right]
\Psi_0^{(1)} (x) + b \left[ \Psi_0^{(1)} (x) \right]^3=0.
\eeq
The corresponding solution $\Psi_0^{(1)} (x)$ must be continuous,
but its first derivative has a jump at $x=0$, meaning that
\beq\label{deriv}
\displaystyle\frac{d \Psi_0^{(1)} (x)}{dx} \biggr\vert \sb{x=+0}
-\frac{d \Psi_0^{(1)} (x)}{dx} \biggr\vert \sb{x=-0}
=-2mu \Psi_0^{(1)} (0)
\eeq
As has been discussed above, two qualitatively different solutions are
possible depending on the sign of $\alpha$, and we must consider
separately two cases.
\end{widetext}

\subsection{Case $\alpha >0$}

In this case, the solution and its first derivative must
decrease exponentially at infinity.
Elementary integration then leads to the result that\cite{Schwabl,Schmischwabl}
\beq\label{sol1}
\Psi_0^{(1)} (x)= \frac{\sqrt{2\alpha}}
{\sqrt{b} \sinh \left[ \sqrt{2m\alpha}|x| +\lambda_1 \right]},
\eeq
where the constant $\lambda_1$, determined from the condition
of Eq.~(\ref{deriv}), is given by
\beq
\lambda_1 ={\rm arctanh} \displaystyle\frac{1}{u}\sqrt{\frac{2\alpha}{m}},
\eeq
Note that the solution Eq.~(\ref{sol1}) makes sense only if $u>0$.
At large $|x|$,
\beq
\Psi_0^{(1)} (x) \approx \displaystyle
\sqrt{\frac{8\alpha}{b}}
\left(
\displaystyle\frac{\sqrt{mu^2}-\sqrt{2\alpha}}
{\sqrt{mu^2}+\sqrt{2\alpha}}
\right)^{1/2} \cdot
e^{-\sqrt{2m \alpha } |x| },
\eeq
in agreement with Eq. (\ref{psigr}), while at $x=0$:
\beq\label{atorig}
\Psi_0^{(1)} (0)= \displaystyle\sqrt{\frac{mu^2 -2\alpha}{b}}
\eeq
This solution is possible only
if $0< 2\alpha <mu^2$. If $2\alpha >mu^2$, the only stable
solution is $\Psi_0^{(1)} (x)=0$. We conclude that
the amount by which $u$ increases the critical temperature is
connected to $\alpha$ by
\beq\label{tshift}
\alpha = \frac{mu^2}{2}.
\eeq
\begin{figure}
\includegraphics[width=6cm, height=7.5cm, angle=-90]{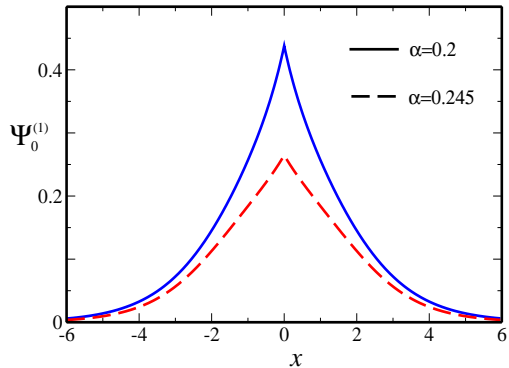}
\caption{ (Color online). Order parameter
$\Psi_0^{(1)}$ plotted as a function of $x$ using Eq.~(\ref{sol1})
for $m=2.0$, $b=1.0$ and $u=0.5$ and two values
of $\alpha$ shown in the figure.
}\label{Figsc1}
\end{figure}

It is instructive to check this result by calculating explicitly the
Green's function form Eq.~(\ref{gfLeq}) and finding its poles.
It is convenient to work in the momentum representation.
Since the problem is translationally invariant in
the $y$ and $z$ directions, we seek solutions of the form
\begin{widetext}
\beq
{\cal G}^{(1)} (X_i =0;\vec r,\vec r^{\prime})=
\displaystyle\frac{1}{(2\pi)^2}\int_{-\infty}^{\infty}\int_{-\infty}^{\infty}
e^{ i(q_y y+ q_z z)} dq_y dq_z \int_{-\infty}^{\infty}\int_{-\infty}^{\infty}
 e^{i(p x +p^{\prime} x^{\prime})}  \;
{\cal G}^{(1)} (\eta_{yz};p,p^{\prime} ) dp dp^{\prime} ,
\eeq
where $\eta_{yz}\equiv \eta (q_y, q_z)=q_y^2 /(2m)+q_z^2 /(2m)$. Substituting
into Eq.~(\ref{gfLeq}) and performing simple manipulations in the
term involving
the $\delta$-function, we obtain two equations to be solved self-consistently
\beq
{\cal G}^{(1)} (\eta_{yz};p,p^{\prime} ) = \displaystyle
\frac{\delta (p+p^{\prime})+u C_1(p^{\prime})}
{\alpha + \eta_{yz}+p^2 /(2m)}, \qquad
C (p^{\prime}) = \int_{-\infty}^{\infty} {\cal G}^{(1)} (\eta_{yz};-
p^{\prime\prime},p^{\prime} )
\displaystyle\frac{dp^{\prime\prime}}{2\pi}
\eeq
The result for the full Green's function is
\beq\label{fullgf}
{\cal G}^{(1)} (\eta_{yz};p,p^{\prime} )=
\displaystyle
\frac{\delta (p+p^{\prime})}{\alpha + \eta_{yz}+p^2 /(2m)}
+\frac{u}{2\pi [1-u K]}\cdot
\frac1{[\alpha +\eta_{yz} +p^2 /(2m)]
[\alpha +\eta_{yz} +(p^{\prime})^2 /(2m)]},
\eeq
\end{widetext}
in terms of the integral
\beq\label{K1}
K \equiv K (\eta_{yz}) &=&\int_{-\infty}^{\infty}\displaystyle
\frac{dp^{\prime\prime}}{2\pi}\cdot
\frac1{\alpha +\eta_{yz} +(p^{\prime\prime})^2 /(2m)}\nonumber\\
 &=& \displaystyle\frac{m}{\sqrt{2m(\alpha+\eta_{yz})}}.
\eeq
The first term in Eq.~(\ref{fullgf}) is the Green's function
corresponding to the absence of any potential, and has the simple
pole at $\alpha=0$.
 The second term is the non-translationally invariant
contribution due to the presence of the defect. For positive $u$ only, the
factor containing $K$ has an additional pole at $\alpha =mu^2 /2$,
implying an increase of the critical temperature
in agreement with Eq.~(\ref{tshift}).

\subsection{Case $\alpha <0$}
In this case, as $|x|\rightarrow\infty$,
$\Psi_0^{(1)} (x)\rightarrow \sqrt{|\alpha|/b}$.
It is easy to complete the integration to obtain\cite{Schwabl,Schmischwabl}
\beq\label{sol2}
\Psi_0^{(1)} (x)=  \left\{
\begin{array}{cl}\
\displaystyle \sqrt{|\alpha|/b} \coth \left[ \sqrt{m|\alpha|}|x|
              + \lambda_2 \right],& u>0, \vspace{.1cm}\\
\displaystyle \sqrt{|\alpha|/b} \tanh \left[ \sqrt{m|\alpha|}|x|
              + \lambda_2 \right],& u<0,
\end{array}\right.
\eeq
\beq\label{lambda2}
\lambda_2= \displaystyle\frac12 {\rm arcsinh}
 \frac{1}{|u|}\sqrt{\frac{4|\alpha|}{m}}.
\eeq
For  $\alpha <0$, potentials with both signs of $u$ lead to physically
sensible positive solutions.
At large distances, $\sqrt{m|\alpha|} |x| \gg 1$,
\beq
\Psi_0^{(1)} (x) \approx
\sqrt{\frac{|\alpha|}{b}}  \left(
1 + \frac{mu \cdot e^{-2\sqrt{m |\alpha| } |x| }}
{\sqrt{|\alpha|}+\sqrt{|\alpha|+mu^2 /4}}
\right),
\eeq
which has the asymptotic form discussed in the previous Section.
Full expressions for $\Psi_0^{(1)} (0)$ can be derived
in a straight forward manner,
but here we only present the less cumbersome ones in the limit
of large and small (with respect to $mu^2$) $|\alpha|$,
\beq
\Psi_0^{(1)} (0) \approx \sqrt{\frac{|\alpha|}{b}} \left(
1+\displaystyle\frac{u}{2}\sqrt{\frac{m}{|\alpha|}}
\right), \quad mu^2 \ll |\alpha|;
\eeq
\beq
\Psi_0^{(1)} (0) \approx  \left\{
\begin{array}{cl}\
\displaystyle\sqrt{\frac{m}{b}}\cdot
\frac{|\alpha|}{m |u|}, & u<0,\vspace{.1cm}\\
\displaystyle\sqrt{\frac{m}{b}} u, & u>0,
\end{array}\right. \quad mu^2 \gg |\alpha|.
\eeq
\begin{figure}
\includegraphics[width=6cm, height=7.5cm, angle=-90]{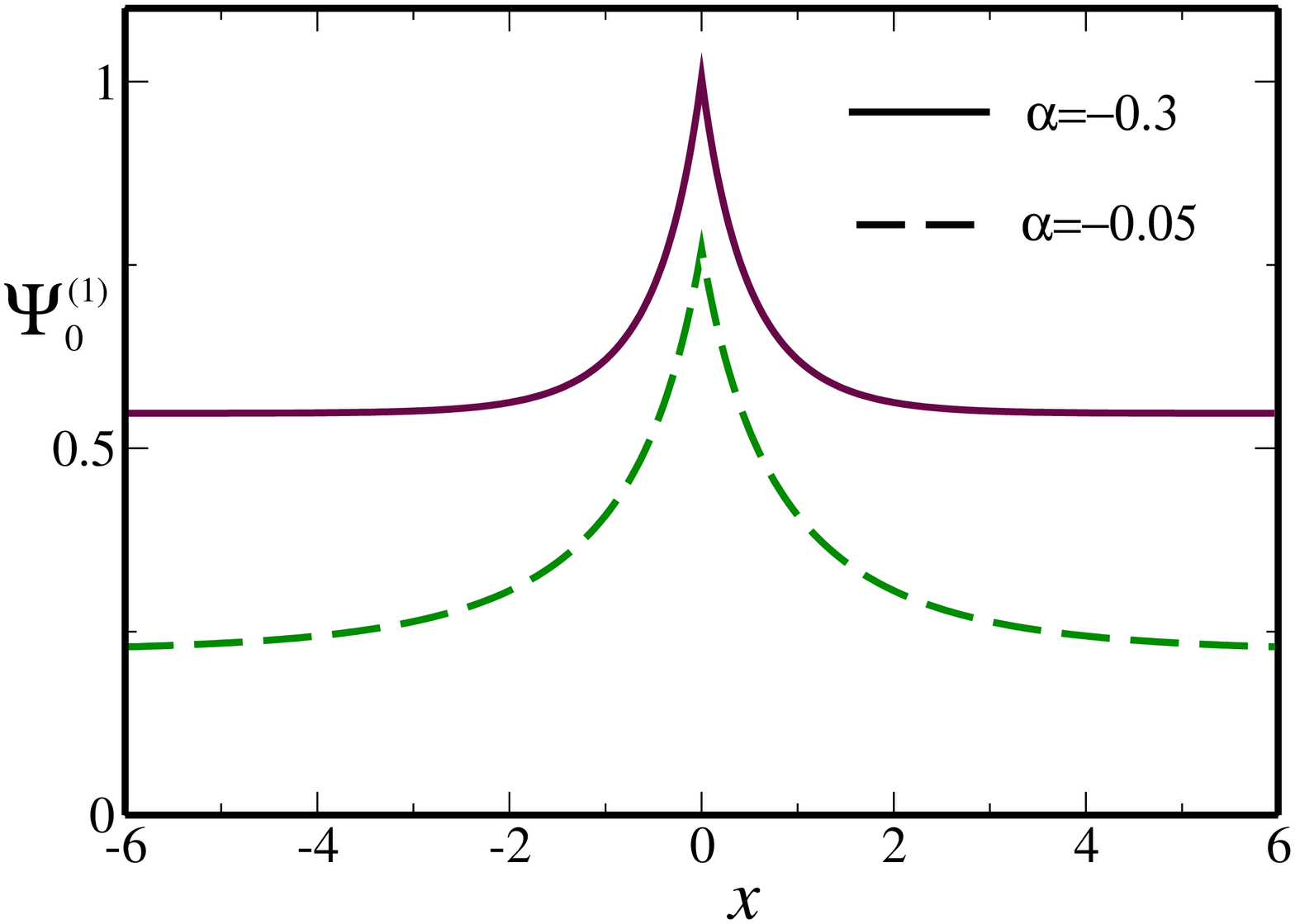}
\includegraphics[width=6cm, height=7.5cm, angle=-90]{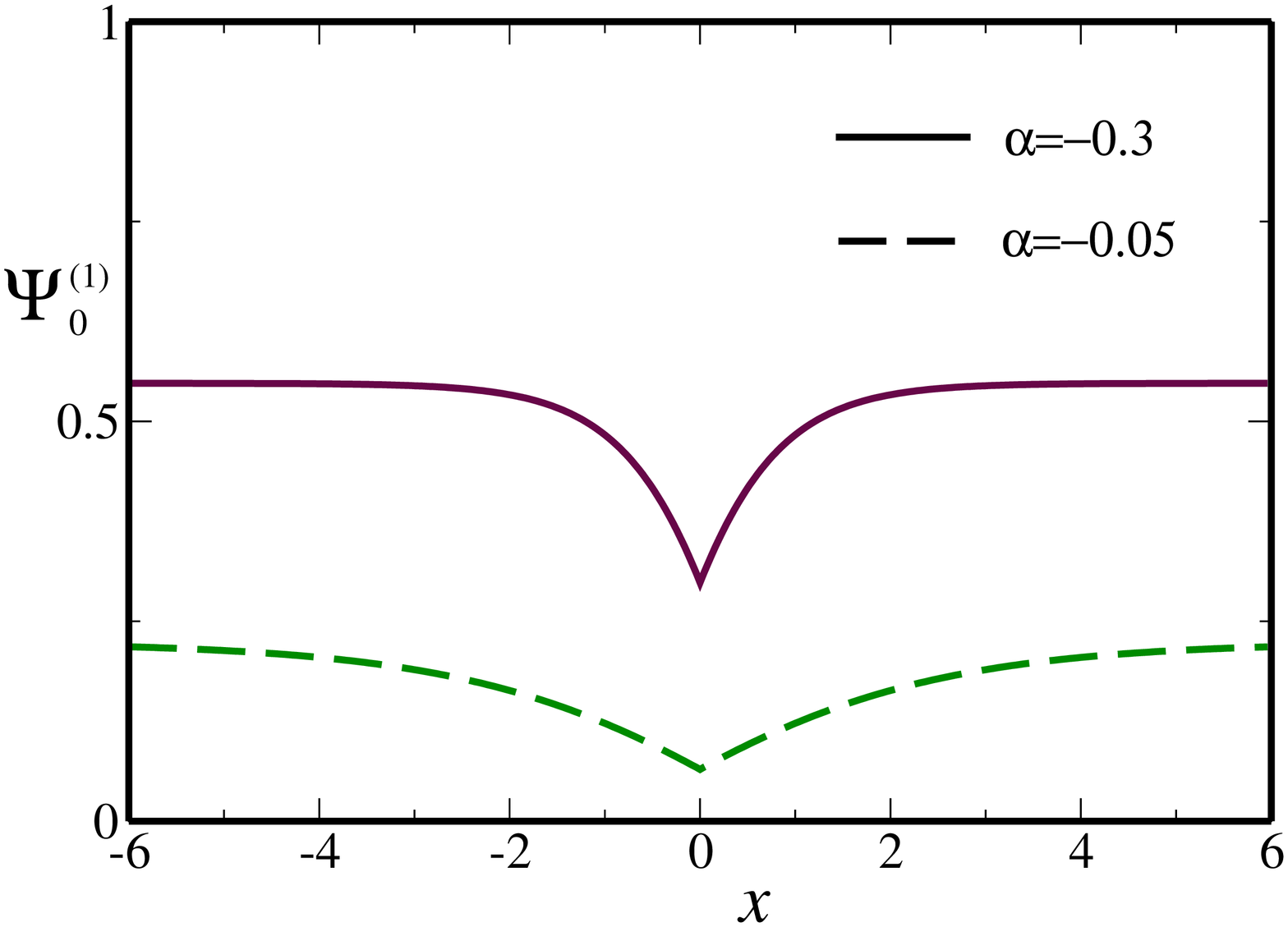}
\caption{ (Color online). Order parameter
$\Psi_0^{(1)}$ plotted as a function of $x$ using Eq.~(\ref{sol2})
for $m=2.0$, $b=1.0$ and $u=0.5$ (upper figure),$u=-0.5$ (lower figure).
The values of $\alpha$ corresponding to each curve are
displayed in the figures.
}\label{Figsc2}
\end{figure}
Depending on the sign of $u$, the order parameter is either greater
or smaller than $\Psi_0^{(0)} (x)= \sqrt{|\alpha|/b}$.
Thus, if $\alpha$ tends to zero from below, the solution for the order
parameter vanishes if $u<0$, but transforms into
\beq
 \Psi_0^{(1)} (x)= \displaystyle\sqrt{\frac{m}{b}}
\frac{u}{mu |x|+1}, \quad \alpha =0,
\eeq
for positive $u$.
The dependence on $x$ becomes a power law, meaning that the influence of the
defect is long range at the special point $\alpha =0$.

We can now substitute the solutions given by
Eqs.~(\ref{sol1}), (\ref{sol2}) into Eq.~(\ref{rho}) and calculate the
average superfluid density $\overline{\rho_s} (T)$.
$u$ is taken to obey the symmetric Gaussian distribution,
\beq\label{gausdis}
 \quad
P[u]= \frac{1}{\sqrt{2\pi}W} \exp \left\{ -\frac{u^2}{2W^2}
\right\},
\eeq
with mean $W$. This form of distribution implies
an exponentially rare probability of occurrence for defects with potentials
with strength much greater than average. However, since $u$ can,
in principle, take any value
in the distribution Eq.~(\ref{gausdis}), the defect-induced enhancement
of the critical temperature is formally unbounded. We must then impose
an upper limit $u_{m}$ on possible values of $u$, which will define the
actual critical temperature
\beq\label{tact}
T_c = T_c^{(0)} \left( 1 + \displaystyle\frac{m u_{m}^2}{2a} \right).
\eeq
The renormalized distribution to be used in Eq.~(\ref{rho}) is
\beq
\Tilde{\mathcal{P}} (\{ u \}) = P[u]/{\mathcal A}, \quad
{\mathcal A}= \int_{-u_{m}}^{u_{m}} P[u] du,
\eeq
and it is simple to integrate over $x$ and obtain
\begin{widetext}
\beq\label{actrho}
\overline{\rho_s} (T) = \left\{
\begin{array}{cl}\
\displaystyle\frac{2  n_i}{b} \int_{\sqrt{2\alpha /m}}^{u_m}
\frac{P[u]}{{\mathcal A}}
\left( u -\sqrt{\frac{2\alpha}{m}} \right) \; du, & \quad
\alpha >0,\vspace{.1cm}\\
\displaystyle\frac{|\alpha|}{b} +\displaystyle\frac{2  n_i}{b}
\sqrt{\frac{|\alpha|}{m}}\int_0^{u_m} \frac{P[u]}{{\mathcal A}}\cdot
\displaystyle\frac{(\coth \lambda_2 -1)^2}{\coth \lambda_2} \; du,
& \quad \alpha<0.
\end{array}\right.
\eeq
\end{widetext}
In deriving the result for $\alpha <0$, we explicitly used
the symmetry, $P[-u]=P[u]$. Noting that, from
Eq.~(\ref{lambda2}),
\beq
\coth \lambda_2 =\left(
\displaystyle\frac{\sqrt{4|\alpha|+mu^2}+\sqrt{m}|u|}
{\sqrt{4|\alpha|+mu^2}-\sqrt{m}|u|}
\right)^{1/2},
\eeq
we see that for $\alpha =0$ the two expressions in Eq.~(\ref{actrho})
become identical. The presence of the $T_c^{(0)}$-enhancing
defects makes the superfluid density finite at that point.
Estimating its order of magnitude, we can write
\beq\label{apprrho}
\overline{\rho_s} (T_c^{(0)}) \sim \displaystyle\frac{n_i W}{b}.
\eeq
The value in the righthand side of Eq.~(\ref{apprrho}) contains the first
power of $n_i$ and is presumed small enough so that the whole approach
based on the Ginzburg-Landau expansion remains valid.

\begin{figure}
\includegraphics[width=7cm, height=8.5cm, angle=-90]{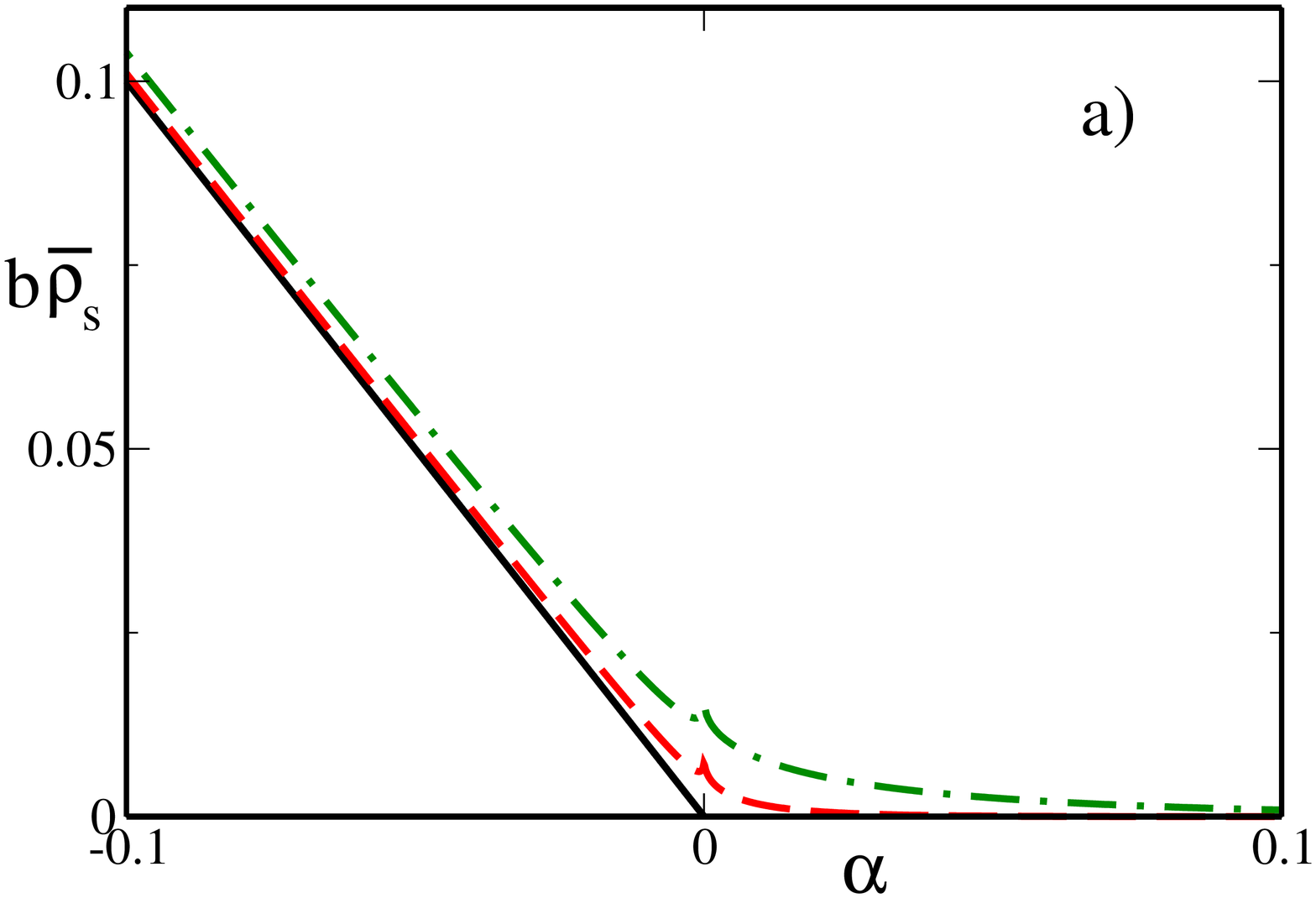}
\includegraphics[width=7cm, height=8.5cm, angle=-90]{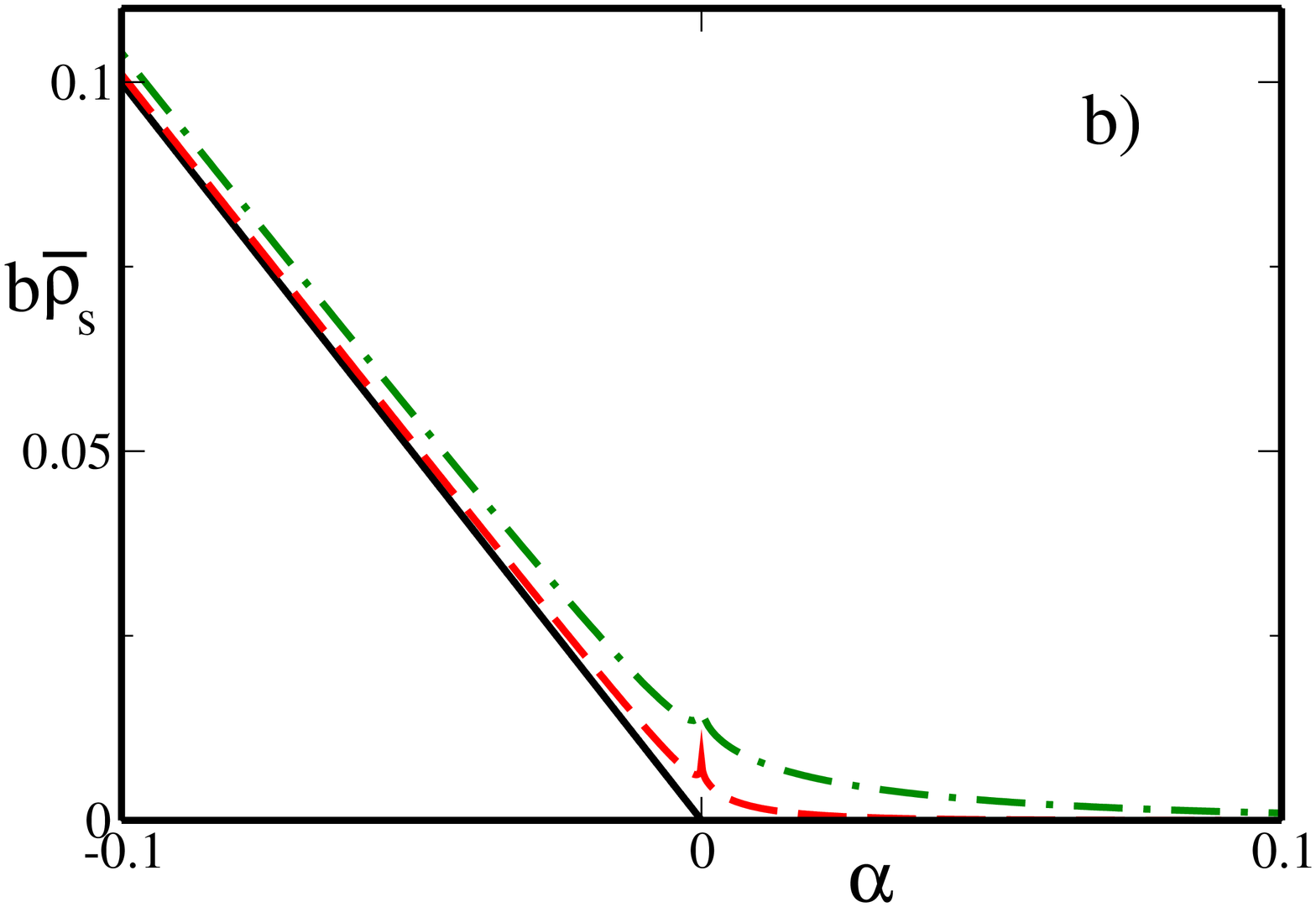}
\caption{ (Color online). The figures show the dependence of
$b\overline{\rho_s} (T)$ as a function of
$\alpha (T) = a ( T -T_c^{(0)})/ T_c^{(0)}$ ($a=1$)
calculated from Eq.~(\ref{actrho}) for the rescaled by means of
Eq.~(\ref{resc}) parameters $n_i=0.1$, $u_{m}=3W$ for figure a)
and $n_i=0.1$, $u_{m}=5W$  for figure b) respectively.
The widths of the disorder distribution
$W=0.0$, $W=0.1$, $W=0.2$ in both figures,
are represented by the solid, dashed and dashed-dotted lines respectively.
}\label{Fig1}
\end{figure}

For convenience, we introduce the rescaled
parameters
\beq\label{resc}
\sqrt{\displaystyle\frac{m}{2}} u \rightarrow u,
\quad
\sqrt{\displaystyle\frac{m}{2}} W
 \rightarrow W, \quad
\sqrt{\displaystyle\frac{2}{m}} n_i \rightarrow n_i \nonumber
\eeq
and plot $b\overline{\rho_s} (T)$ as a function of
$\alpha (T) = a ( T -T_c^{(0)})/ T_c^{(0)}$, assuming $a=1$.
The results for several values of disorder distribution width $W$
( $W=0.0$,  $W=0.1$, and $W=0.2$ ) and $u_{m}=3W$ on all plots,
are presented in Fig.~\ref{Fig1}.

From the plots it follows that, if the values of $u$ are broadly
distributed and $u_m \gg W$ belongs to the Lifshitz
tail, the superfluid density for larger $\alpha$ is exponentially small.
Under the same circumstances, the behavior near the point $T=T_c^{(0)}$ is
not sensitive to the exact value of $u_m$.
We note also that for $u_m \gg W$, ${\mathcal A}\approx 1$ and is not
of much importance. Below $T_c^{(0)}$, however, the behavior asymptotically
approaches that of the disorder-free system.
These results are in qualitative agreement with the solution for the
order parameter presented in Ref.~\onlinecite{Schwabl}, having
a small but finite value going to zero at some weakly
$n_i$-dependent value $T_c > T_c^{(0)}$.
\begin{figure}

\includegraphics[trim=-1cm -1cm 0 0, width=83mm, height=9cm]{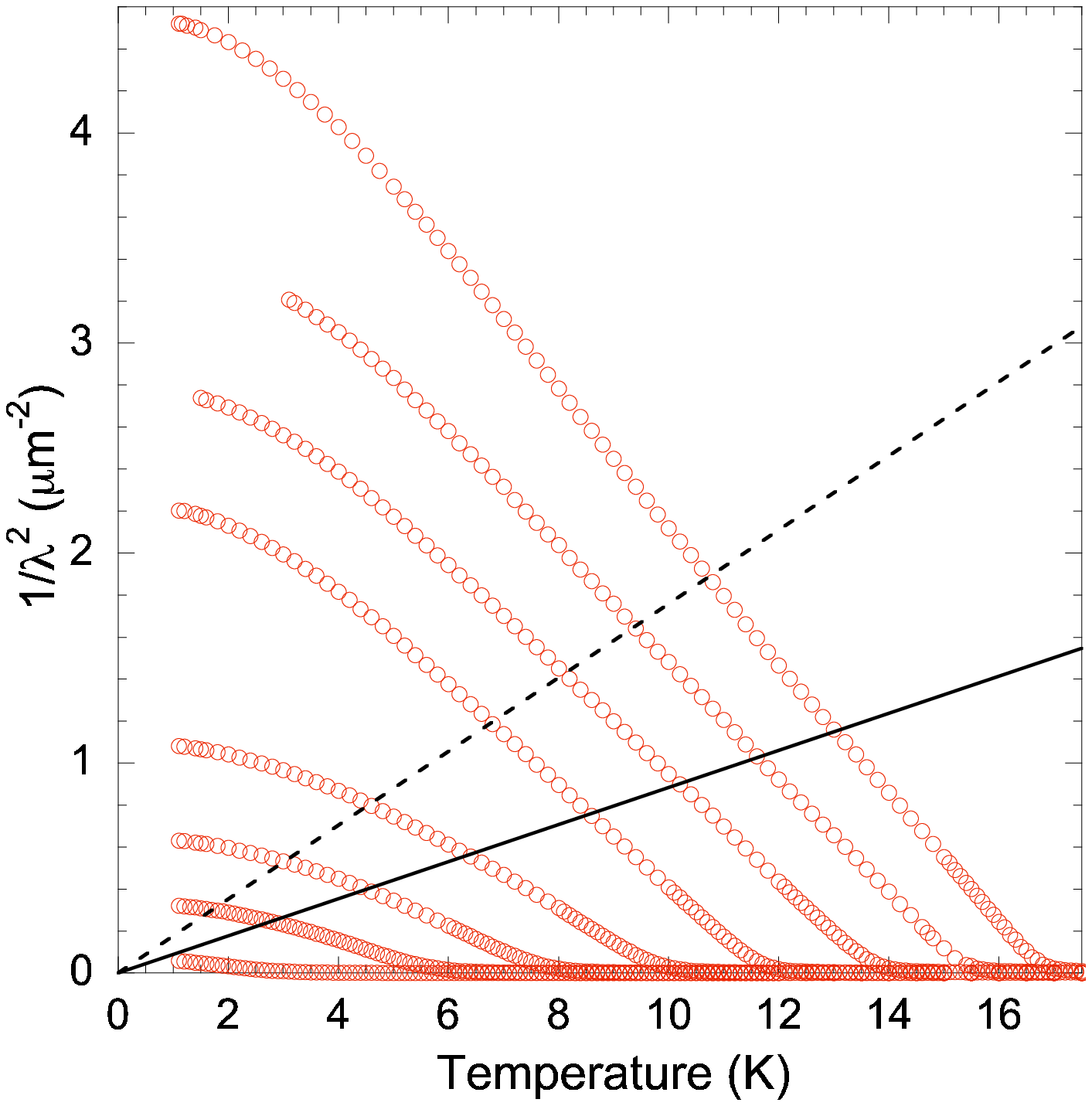}
\vspace{1mm}
\includegraphics[width=83mm, height=6cm]{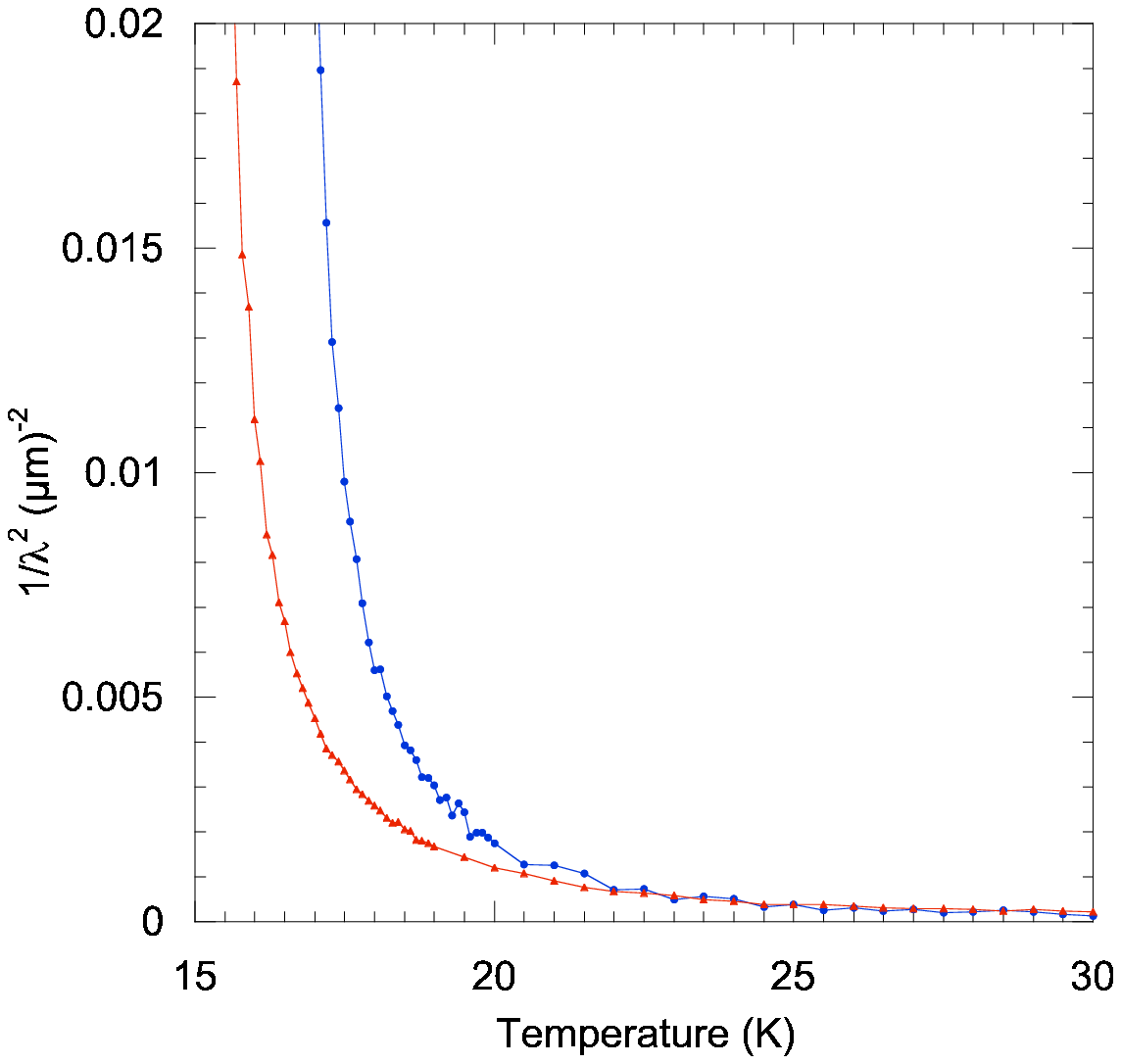}
\caption{Measured superfluid density in the underdoped regime of
${\rm Y} {\rm Ba}_2 {\rm Cu}_3 {\rm O}_{6.333}$ taken from
Ref. ~\onlinecite{Broun}. The data suggest the presence of
decreasing superconducting response above some temperature.
The meaning of the curves is explained in the text.
}\label{Fig2}
\end{figure}

In Fig.~\ref{Fig2}, we present data for
experimental measurements of the superfluid density taken from
Ref.~\onlinecite{Broun}. The upper figure shows the results for the
superfluid density of an ${\rm Y} {\rm Ba}_2 {\rm Cu}_3 {\rm O}_{6.333}$
ellipsoid, measured as a function of temperature for different values
of doping starting from the most ordered sample with $T_c =17K$.
The lower figure describes the behavior of $\rho_s (T)$ very close to
the critical temperature for two levels of doping that lead to the
highest $T_c$'s ( $\approx 17K$ and $\approx 15K$ ) as shown in the
upper figure. In experiments described in Ref.~\onlinecite{Broun},
the superfluid density
was determined from the penetration depth measurements done on the high
quality crystals. Highly ordered samples were prepared by extended
annealing under pure oxygen gas flow, as well as high hydrostatic
pressure at room temperature to enhance CuO-chain
ordering\cite{Liang1,Liang2}. This procedure removes randomness in the
spatial distribution of isolated oxygen vacancies that could act as pinning
and scattering centers. However, planar-type inhomogeneities
are present as twin boundaries as these materials are not detwinned.
In conventional superconductors, the twin boundaries are well known to
increase locally the critical temperature\cite{khlyust}.
In ${\rm Y} {\rm Ba}_2 {\rm Cu}_3 {\rm O}_{6.333}$,
the twin boundaries parallel to the $c$-axis
are envisioned as the planes separating two regions in which the
CuO chains are oriented perpendicular to each other, and
are likely to serve as a source of $T_c$-enhancement as well.
The reason for this is that it is advantageous for the
oxygen vacancies to be located near the twin
boundaries\cite{Durst}. And the presence of narrow regions with surplus
of these vacancies means the enhancement of superconductivity,
and as a consequence higher critical temperature in the vicinity.
For each plane, this is effectively modeled by adding the potential
Eq.~(\ref{ourpot}) to the conventional GL action. Randomness of $u$
is likely to come from variation of the in-plane
concentration of vacancies, stemming in its turn from the non-uniformity
of the initial density of the twin boundaries. Since, in our case all
such planar defects appear to increase the critical temperature,
the values of $u$ are determined relative to some average value, hence
having both signs. $T_c^{(0)}$ defined previously as the critical
temperature of completely disorder-free sample must thus be regarded
shifted up so that Eq.~(\ref{disconstr}) is satisfied.
We should mention also that the issue of local superconductivity enhancement
due to the twin-boundaries themselves has been considered previously
in Ref.~\onlinecite{Abrikosov}. The twinning planes were assumed
to form a periodic array and be described in the GL functional
by the sum of terms of the form Eq. (\ref{imppot}) all having the same $u$.
In our approach, we specifically highlight the importance of
randomness in potentials for the purpose of the qualitative interpretation
of the measured superfluid density.

It is seen from the data in Fig.~\ref{Fig2} that a small but
finite superfluidity persists above some temperature obtained
by extrapolating the straight lines, describing the behavior at
lower $T$ down to zero.
If we associate this value with $T_c^{(0)}$ in our approach,  we can
claim that the simple model of planar disorder reproduces
qualitatively the experimentally observed temperature
dependence of $\rho_s$. From the data, though, it is difficult to
infer the precise value of the actual critical temperature $T_c$.
Since the aim of this calculation is mainly to
illustrate the generic features resulting from randomness in $T_c$'s,
we have not attempted to determine the choices of parameters $W$, $u_m$,
$a$ required for a precise match between the theoretical and
experimental curves.
Already this qualitative agreement suggests that even
in experiments done on high quality samples, there are local regions with
critical temperatures significantly above the average $T_c$ . The
smallness of superfluid density in the tails suggests
that such regions are rare and have broadly distributed local critical
temperatures.

\section{Discussion and Conclusions}

In this Section, we discuss the relevance of the simple model
of random planar defects to the experimentally observed behavior of
the superfluid density as a function of temperature close to $T_c$.

First, it is appropriate to ask what changes to the behavior of
$\rho_s$ are expected if we go to the next orders of expansion in defect
concentration. To examine this, consider how one proceeds to
obtain the correction to the superfluid density that is of second
order in $n_i$. According to the general strategy, it
is necessary to solve the equation for the order parameter
in the presence of two defects located at points $X_1$ and $X_2$.
The distance $L=|X_2-X_1|$ emerges as a new parameter for the problem,
and together with the strengths of the potentials will determine the point
at which the real positive solution $\Psi_0^{(2)} (L;x)$ first obtains.
To find it, one can easily solve the corresponding
generalization of Eq.~(\ref{gfLeq}) for the Green's function
${\cal G}^{(2)} (X_1 =0, X_2 =L;\vec r,\vec r^{\prime})$ of the operator
${\hat L}_{0} [u]$ with the potential
\beq\label{npot}
u (\vec r)=-u_1 \delta (x) -u_2 \delta (x-L).
\eeq
The subscript $(2)$ means now that we do all calculations in
the presence of exactly two defects.
The solution has the form
\beq\label{gf2sol}
& {\cal G}^{(2)} (\eta_{yz};p,p^{\prime} ) = & \nonumber \\
&\displaystyle
\frac{\delta (p+p^{\prime})+u_1 C_{1}( p^{\prime})+
u_2 C_{2}(p^{\prime}) e^{-ip L}}
{\alpha + \eta_{yz}+p^2 /(2m)},&
\eeq
\vspace{1mm}
\beq
C_{1} (p^{\prime}) = \int_{-\infty}^{\infty} {\cal G}^{(2)}
(\eta_{yz};-p^{\prime\prime},p^{\prime} )
\displaystyle\frac{dp^{\prime\prime}}{2\pi},
\eeq
\beq
C_{2} (p^{\prime}) = \int_{-\infty}^{\infty} e^{-i p^{\prime\prime} L}
{\cal G}^{(2)} (\eta_{yz};-p^{\prime\prime},p^{\prime} )
\displaystyle\frac{dp^{\prime\prime}}{2\pi}.
\eeq
We will not write out in full the cumbersome expression for the Green's
function
and state only that the poles of ${\cal G}^{(2)} (\eta_{yz};p,p^{\prime})$,
indicating the occurrence of a transition at some $T> T_c^{(0)}$,
are determined from the equation
\beq\label{det}
\left( 1- u_1 \sqrt{\displaystyle\frac{m}{2\alpha}} \right)
\left( 1- u_2 \sqrt{\displaystyle\frac{m}{2\alpha}} \right)- \nonumber\\
\displaystyle\frac{m u_1 u_2}{2\alpha}
\cdot e^{-2\sqrt{2m\alpha} |L|} =0.
\eeq
The left hand side of Eq.~(\ref{det}) is just the
corresponding determinant (taken at $\eta_{yz}=0$)
which arises in the process of solving the system of two
coupled linear equations for $C_{1} (p^{\prime})$ and $C_{2} (p^{\prime})$.
If $u_1 = u_2$, Eq.~(\ref{det}) reduces (upon the proper rescaling)
to the known result\cite{Schmischwabl}.
It is clear from Eq.~(\ref{det}), that for positive $u_1$ and $u_2$,
the critical value of $\alpha =(m/2) \max(u_1^2, u_2^2)$ if $|L|=\infty$,
implying a complete independence of the defects. But for $|L|=0$,
the pole occurs at $\alpha =(m/2)(u_1+u_2)^2$, in agreement with the fact
that if both defects are located at one point, their strengths
simply add. It follows then that placing the second defect with
positive $u$ at any finite distance $L$, in addition to the
one already present, increases the critical $\alpha$.
This means that if there are two defects located not too
far from each other in the sample, with strengths close to $u_m$,
the actual critical temperature, as a result of going to the second
order in $n_i$, will be higher than that given by Eq. (\ref{tact}).
However, this circumstance does not affect the qualitative
interpretation of the data since, as was mentioned
before, the presumed exponential smallness of $\rho_s (T)$
at temperatures considerably higher than $T_c^{(0)}$, renders the
actual $T_c$ difficult to determine from Fig.~\ref{Fig2}.
Thus we will not discuss further all the
calculations to second and higher orders in $n_i$,
but rather note the following. Once the potential strengths
are broadly distributed, the value of $b\overline{\rho_s} (T)$
at $T=T_c^{(0)}$ is not sensitive to the actual $T_c$.  To
calculate it one can safely set $u_m =\infty$.
However, the calculated $b\overline{\rho_s} (T)$ using Eq.~(\ref{actrho})
has a peak at $\alpha =0$ as seen in Fig.~\ref{Fig1}.
We believe that this non-monotonic behavior in the vicinity of
$T_c^{(0)}$, is an artifact of the mean-field approximation used
in our treatment from the very beginning. Thermal fluctuations
when $\alpha$ is small, are expected to renormalize down the values of
the superfluid density, but consideration of this question is beyond
the scope of this paper.

The planar defects considered in this paper, may not
be the only ones present despite the high quality of the
samples. Isolated and rare point defects due to oxygen
disorder, other lattice defects such as dislocations\cite{Nabutovskii} are,
in principle, not excluded\cite{Berlinsky}.  It is important, however,
that among all possible types of defects, the plane-like ones ensure
the broadest possible distribution of the local critical temperatures.
This follows from the generalized Harris criterion\cite{Harris, Wallin},
arguing that whenever $2-d^{*}\nu >0$, with $\nu$ being the correlation
length critical exponent and $d^{*}$ the number of dimensions in which the
system is random, disorder is relevant near the critical point.
For point and columnar disorder, $d^{*}$ is equal to $3$ and $2$
respectively, whereas in the case of stacked planar defects $d^{*}=1$.
Obviously, the greater $2-d^{*}\nu$ is, the wider the distribution of local
$T_c$'s is due to randomness in parameters
characterizing the defect potential\cite{tvojta}.  One should remember, though,
that not all defects, but only the extended ones of sizes greater than
$\xi_0$, can be satisfactorily accounted for within the framework of
GL theory.  Others should be treated using a suitable
microscopic model.  If necessary, possibly in lower quality samples,
defects with spherical and cylindrical shapes
must be considered as well, but these are likely to
affect the superfluid density in a much narrower region
around $T_c^{(0)}$.  Investigation of such defects is left for future work.

\begin{acknowledgments}
The authors would like to thank Prof. David Broun for many helpful conversations and for providing the data presented in the figures above. CK and AJB are supported by the Natural Sciences and Engineering Research Council of Canada, by the Canadian Foundation for Innovation and by the Canadian Institute for Advanced Research.
\end{acknowledgments}

\begin{widetext}
\appendix
\section{Expansion in powers of $n_i$ }

In this Appendix, we obtain the average of the function
$F_N ( \{ u_1 \}\vec R_1, \ldots ,\{ u_N \}\vec R_N ;\vec r)$
given by Eq. (\ref{avgen}) in the form of a series in powers of
defect concentration $n_i$. The corresponding derivation is straightforward
and follows the lines of Ref.~\onlinecite{Lifshitz}.
We should notice first, however, that the function in angular brackets
in Eq.~(\ref{avgen}) is not symmetric,
but the final result will not change if we replace
$F_N ( \{ u_1 \}\vec R_1, \ldots ,\{ u_N \}\vec R_N ;\vec r)$
with the function
\beq
F_{sN} ( \{ u_1 \}, \ldots ,\{ u_N \}; \vec R_1, \ldots ,\vec R_N ;\vec r )=
\frac{1}{N!} {\hat P} \left[ \{ u_1 \},\ldots ,\{ u_N \} \right]
F_N ( \{ u_1 \}\vec R_1, \ldots ,\{ u_N \}\vec R_N ;\vec r),
\eeq
symmetrized over all sets $\{ u_i \}$ for a particular
location of the defect.( ${\hat P}$ is the symmetrization
operator) This creates functions symmetric with respect
to permutations of coordinates $\vec R_1, \ldots ,\vec R_N$.
It is possible to verify then that for any finite number of such functions
$F_{sm} (\{ u_1 \}, \ldots ,\{ u_m \}; \vec R_1, \ldots ,\vec R_m ;\vec r)$,
($m\ge 0$)
\begin{multline}\label{genfor}
F_{sN} ( \{ u_1 \}, \ldots ,\{ u_N \}; \vec R_1, \ldots ,\vec R_N ;\vec r )=
F_0 (\vec r) +\sum_{i} \Phi_{1} (\{ u_i \};\vec R_i;\vec r)
\\
+\sum_{i<j} \bigl[ \Phi_{s2} (\{ u_i \},\{ u_j \};\vec R_i,\vec R_j;\vec r)-
\Phi_{1} (\{ u_i \};\vec R_i;\vec r)-
\Phi_{1} (\{ u_j \};\vec R_j;\vec r) \bigr]
\\
+\sum_{i<j<k}
\bigl[ \Phi_{s3}
(\{ u_i \},\{ u_j \},\{ u_k \};\vec R_i,\vec R_j,\vec R_k;\vec r)
-
 \Phi_{s2} (\{ u_i \},\{ u_j \};\vec R_i,\vec R_j;\vec r)
- \Phi_{s2} (\{ u_i \},\{ u_k \};\vec R_i,\vec R_k;\vec r)
\\
-\Phi_{s2} (\{ u_j \},\{ u_k \};\vec R_j,\vec R_k;\vec r)
+ \Phi_{1} (\{ u_i \};\vec R_i;\vec r)+
\Phi_{1} (\{ u_j \};\vec R_j;\vec r)+ \Phi_{1} (\{ u_k \};\vec R_k;\vec r)
\bigr]
+ \cdots
\end{multline}
In Eq. (\ref{genfor}) ($m\ge 2$),
\beq\label{Phis}
\Phi_{sm} ( \{ u_1 \}, \ldots ,\{ u_m \}; \vec R_1, \ldots ,\vec R_m ;\vec r )
 = F_{sm} ( \{ u_1 \}, \ldots ,\{ u_m \}; \vec R_1, \ldots ,\vec R_m ;\vec r )
-F_0 (\vec r),
\eeq
and these
$\Phi_{sm} ( \{ u_1 \}, \ldots ,\{ u_m \}; \vec R_1,\ldots ,\vec R_m ;\vec r )$
are also symmetric. $F_0 (\vec r)$ is a function of $\vec r$ only,
and it, as well as $\Phi_{1} (\{ u_i \};\vec R_i;\vec r)$, does not require
symmetrization.
Assuming that all defects are located at points $\vec R_i$, and that
$F_0 (\vec r)$ is the value in the absence of defects, we can take
the thermodynamic limit $N\rightarrow\infty$, $V\rightarrow\infty$,
$n_i =N/V$
in Eq.~(\ref{genfor}) to get
\begin{eqnarray}\label{spav}
&\bigl\langle F_N ( \{ u_1 \} \vec R_1, \ldots ,\{ u_N \} \vec R_N ;\vec r )
\bigr\rangle = F_0 (\vec r) +
n_i \int \Phi_{1} (\{ u_1 \};\vec R_1;\vec r) d\vec R_1 & \nonumber
\\
&+\displaystyle\frac{n_i^2}{2!} \iint \bigl[
\Phi_{s2} (\{ u_1 \},\{ u_2 \};\vec R_1,\vec R_2;\vec r)-
\Phi_{1} (\{ u_1 \};\vec R_1;\vec r) - \Phi_{1} (\{ u_2 \};\vec R_2;\vec r)
\bigr]
w_2 (\{ u_1 \},\{ u_2 \};\vec R_2-\vec R_1) \: d\vec R_1 d\vec R_2  & \nonumber
\\
&+\displaystyle\frac{n_i^3}{3!} \iiint \bigl[
\Phi_{s3} (\{ u_1 \},\{ u_2 \},\{ u_3 \};\vec R_1,\vec R_2,\vec R_3;\vec r)
-\Phi_{s2} (\{ u_1 \},\{ u_2 \};\vec R_1,\vec R_2;\vec r)
-\Phi_{s2} (\{ u_2 \},\{ u_3 \};\vec R_2,\vec R_3;\vec r) &  \nonumber
\\
& -\Phi_{s2} (\{ u_3 \},\{ u_1 \};\vec R_3,\vec R_1;\vec r)
+\Phi_{1} (\{ u_1 \};\vec R_1;\vec r) +\Phi_{1} (\{ u_2 \};\vec R_2;\vec r)
+\Phi_{1} (\{ u_3 \};\vec R_3;\vec r)
\bigr] \cdot & \nonumber
\\
 & w_3 (\{ u_1 \},\{ u_2 \},\{ u_3 \};\vec R_2 -\vec R_1, \vec R_3 -\vec R_2 )
 \: d \vec R_1 d \vec R_2 d \vec R_3 + \cdots &
\end{eqnarray}
The full average then follows in a straightforward way by substituting
this average over positions into Eq.~(\ref{avgen}).
In Eq.~(\ref{spav}), $w_2 (\{ u_1 \},\{ u_2 \};\vec R_2-\vec R_1)$
is the probability of the simultaneous occurrence of two defects
at points $\vec R_1$ and $\vec R_2$ characterized by the
sets $\{ u_1 \}$ and $\{ u_2 \}$ respectively.
Obviously, $w_2 (\{ u_1 \},\{ u_2 \};\vec R_2-\vec R_1)=
w_2 (\{ u_1 \},\{ u_2 \};\vec R_1-\vec R_2)$.
Probabilities $w_3, w_4,\ldots$ in higher order terms have
analogous meaning, and in general depend on $\{ u_i \}$. However, this
dependence is likely to be noticeable only if defects are close
to each other.
All functions $w_m$ will depend on concentration
$n_i$, but in the limit all $|\vec R_i-\vec R_k| \gg 1/(n_i)^{1/d}$,
they quickly tend to unity. Those functions must also factorize if one
of the coordinates tends to infinity, meaning, for example, that
$w_3 (\{ u_1 \},\{ u_2 \},\{ u_3 \};\vec R_2 -\vec R_1, \vec R_3 -\vec R_2 )
\rightarrow w_2 (\{ u_1 \},\{ u_3 \};\vec R_3-\vec R_1)$,
if $|\vec R_2| \rightarrow \infty$.
\end{widetext}

\end{document}